# Virtualization of Electromagnetic Operations (VEMO)

Steven Jones
Johns Hopkins University Applied Physics Laboratory
Laurel, Maryland, USA

David Coleman
Johns Hopkins University Applied Physics Laboratory
Laurel, Maryland, USA

Robert Nichols
Johns Hopkins University Applied Physics Laboratory
Laurel, Maryland, USA

*Abstract*—Today's operations in the spectrum occur across many disparate and unique devices. For example, a military or commercial maritime platform can have dozens of apertures used in various functions. Utilizing software defined radios and dynamic analog front ends, we propose that future systems involving multiple applications of RF, virtualize operations to gain performance efficiencies. The concept of virtualization of electromagnetic operations (VEMO) is considered a means to enable flexible use of hardware and spectrum resources to achieve diverse mission requirements in military and commercial settings. This will result in more efficient use of resources, including size/weight/power and integration of objectives across systems that employ the electromagnetic spectrum.

*Keywords—spectrum, antenna, software-defined radio, cognitive radio, resource allocation*

## I. Introduction

In terms of space, time, and frequency the electromagnetic spectrum is vast; however, it is a limited resource for practical applications. Demand for bandwidth in radio frequencies (RF) is increasing significantly in the commercial sector due to the need to sense and communicate over large regions and in population centers. Military applications within the RF spectrum include radar, communication and network systems, remote control of unmanned systems, and electronic warfare – and various subgroups within these broad categories. As such, RF spectrum usage fluctuates widely across regions, times of day, and even during missions. Moreover, the nature of devices and propagation limits the desirability of frequencies according to the constraints of mobility and antenna coverage.

Today RF spectrum is assigned in a very static and antiquated manner that is not efficient, but that protects access for approved purposes. Most of the electromagnetic spectrum is allocated exclusively to a certain user or licensee in a given location for all time. This leads to inefficient spectrum utilization since spectrum may be reserved but unused at any given time.

Even though significant research has been executed in the area of spectrum sharing protocols over the past 20 years that promises to increase spectrum efficiency (in time), such as dynamic spectrum access (DSA) and the Citizen Broadcast Radio Service, spectrum efficient applications are not being deployed for military systems within their exclusive bands.

The size, weight, and power (SWaP) of military platforms are also changing – from large monolithic systems to a variety of small systems distributed across the battlefield. The shift to small platforms restricts the number of antennae present on a system yet the capabilities required to survive on the battlefield remains high. This means the small platform must do more with less hardware than its predecessor did. Furthermore, given the dynamic nature of the modern battlefield, it may be advantageous to shift mission functions amongst the distributed platforms than to design new (smaller yet still) single function systems (e.g., jam from the drone closest to the target and not necessarily from any one particular drone). Some of these concepts are also seen in the commercial sector with many small drone concepts that are used for security, logistics and transportation.

A new paradigm in RF system development is required to address inefficient spectrum utilization and distributed electromagnetic operations on the battlefield – the virtualization of electromagnetic operations (VEMO). This concept extends the innovations of virtualizations from the communications domain with software-defined radios and dynamic spectrum techniques in communications, radar and sensing. By hosting these functions on a virtualized platform, a wide range of optimization and efficiencies will be available that cross applications. These advantages come along with the existing benefits of software-defined techniques like flexibility and adaptability in implementations.

## II. Related Research

While the concept of software-defined radio is typically associated with communications, the use of a software-defined paradigm has grown beyond communications [1] to include radar [2], aircraft and drone detection [3], medical imaging [4], and navigation. [5, 6]. Each application uniquely employs the software-defined constructs with many of the same benefits of SDRs (e.g., reuse, multi-purpose) and a similar diversity of underlying hardware platforms to include FPGAs and GPUs. Moreover, the notion of cognitive capabilities, where the system has varying degrees of autonomous behavior, has come up in both communications and radar applications [7]. Virtualized implementation of the cognitive engine and radar processes enables more computational resources and the ability to adapt processes during the mission.

While the use of software-defined approaches is expanding, combined functionality – as is proposed herein – is still an emerging concept. The combination of radar and communication has been demonstrated in [8] and [9] where each (separately) achieved dual applications in the transceiver through engineered waveform design – not an automated process; and this concept has not been extended for military utility in other areas (e.g., electronic warfare, navigation). For this reason, there is potential to virtualized operations across all applications through the extensive and complete change of system-specific RF systems.

For the past twenty years the DoD has supported programs intended to converge the use of apertures, especially in large surface ships. These have included the Advanced Multifunction RF Concept (AMRFC) [12], Integrated Topside (INTOP) [13], the Surface Electronic Improvement Program (SEWIP) [14], and Arrays at Commercial Timescales (ACT) [15]. Note that while these efforts have focused on aperture and RF component sharing, they have not gone as far as sharing waveforms as we propose herein. AMRFC was an ONR program starting in 1998 sharing amongst receivers and transmitters on a single electronically steered array. The focus was on analog beamforming, although some studies included digital arrays. The INTOP program was a follow-on effort to spin off separate capabilities into programs of record, including those employing EW, communications, and radar. This included partly transitioning into SEWIP. More recently, the DARPA ACT program has been developing a common module for core functionality in an array. This will enable reconfigurable and tunable apertures.

Numerous academic studies on joint application designs, including shared waveforms have emerged. Most of these address the combination of pairs of applications such as radar and communications, sensing and communications, EW and radar, EW and comms, and PNT and comms. VEMO takes this collaboration to it natural extreme by combing all of the above.

## III. VEMO CONCEPT

### A. Overview

Wireless systems coexist though manual spectrum coordination to avoid interference in the presence of other applications. Automated techniques including dynamic spectrum access in unlicensed spectrum demonstrate one method to improve spectral efficiency, but DSA is still limited to singular functions such as communications. Combining automated techniques and measurements (e.g., spectrum monitoring, channel state) across applications such as communications and radar systems, the spectral efficiency is increased. In the VEMO concept, the waveforms and signal processing of applications are virtualized to achieve full cooperation and coordination across spectrum operations.

With VEMO, collaboration and coordination will take place from baseband processing through the aperture and into the spectrum utilization. This opens up new opportunities to share antenna arrays, signal processing algorithms, and baseband processing between communications, radar, and electronic warfare (e.g., reactive jammer). For example, communication and sensing missions may utilize a common waveform to concurrently conduct radar and jamming missions rather than transmitting separate signals. A position, navigation, and tracking (PNT) system receiver could feed its received information into a signals intelligence (SIGINT) system operating within a common bandwidth and frequency. Further optimizations (in time, space, frequency, and mission) across architecture and control plane may be available in VEMO verses traditionally stove-piped systems with dedicated functionality.

Target spectrum applications for VEMO include radar, surveillance, communications, electronic warfare (EW), cyber, PNT, and military information support operations (MISO). To ensure mission efficiency is not degraded using VEMO, requirements spanning all systems must be understood including but not limited to: antenna performance, dynamic range of the RF front end, sampling rate, digital signal processing cycles available, saturation limits, and a common lexicon amongst applications. VEMO must also operate under a set of policies that dictate limitations due to maneuver and administrative actions, rules of engagement, mission prioritization and system de-confliction, and various configurations. Fig 1 illustrates the functional components of VEMO.

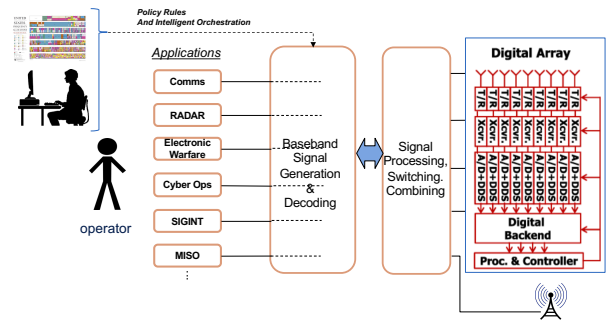

Fig. 1. VEMO Concept Block Diagram

### B. Challenges and Opportunities

There are numerous implementation challenges within VEMO, within each application and across the set of missions. For instance, there are numerous communication systems on any military platform (e.g. satellite, UAV, vehicle, ship) that have differing requirements, antennas, and signal processing, while using different spectrum bands. A platform will have multiple radars, with differing requirements and purposes. Initiatives noted in early cited references have been made to employ versatile hardware such as software defined radios that can host a variety of waveform processing that would allow the user to select the signals to be used at any given time on that hardware. Each system has its own requirements and devices are designed to meet those requirements expressly. There is not typically excess processing capacity or spectral coverage that will allow a system to meet the needs of other systems due to a lack of funding and business incentive.

The VEMO concept addresses a union of requirements across numerous current systems – not individual stove-piped systems - and facilitates greater capabilities in size, weight, and power constrained platforms. The information sharing facilitated by VEMO will allow surges by one application into the collective set of resources based on priorities, thereby improving overall mission performance. With that said, significant challenges exist related to dynamic range of RF applications (e.g., high power of electronic attack while simultaneously listening for signals in electronic and signals intelligence). VEMO would also need to balance the scheduling and resource requirements. For example, the integration period for the radar application may exceed the coherence time for a communication application thus limiting the overall throughput. Therefore, there may need to be engineering compromises – or a systems of systems approach - in particular applications to gain the overall benefits.

*C. VEMO Enables New Applications*

VEMO enables joint sensing and communication applications with respect to temporally and spatially separate actions. For example, one challenge in directional networking is neighbor discovery. If a radar application shared its data with the network controller, neighbors are discovered using sensing verses through announcements – reducing network overhead and network RF signature. Furthermore, radar applications can extract the return signals from within the received communication signal using interference cancellation techniques and improve the dynamic range by demodulating the strong communications signal, regenerating it and then subtracting from the bulk signal to obtain the radar return signal. VEMO also enables multi-input, multi-output radar using both radar and communication signals [10]. Further ideas and survey on the concept of joint radar and communication have been recently published [9].

For electronic warfare in a military setting, the coordination of communications with jamming is paramount. Jamming signals jointly developed with communications requirements can aid in cancellation and interference avoidance while simultaneously connecting disparate devices – thereby reducing a complex mission-deconfliction process. Likewise, multi-static radar could facilitate passive platforms which have low probability of detection advantages. The platform would not need to transmit a radar signal, but rather would collect returns from other radar systems and signals of opportunity. By having these functions jointly executed in VEMO, there would be more efficiency in control and coordination.

One could argue that multiple systems can perform these functions today. However, the interfaces between modern systems do not support collaborated and coordinated functions in addition to the precise timing requirements. For these reasons, VEMO would dramatically change our ability to execute these combined functions in the electromagnetic spectrum.

IV. VEMO DESIGN

*A. Requirements*

Understanding the requirements that drive the design of each wireless defense system is required. Available bandwidths, propagation, threats, cost, and availability of devices to achieve transmit powers and ranges that are required are drivers that have led to the frequency assignments of each system today. Those requirements will not change. However, by sharing capabilities among systems, it may be possible to exceed those requirements. Characterizing the performance criteria of each elemental defense system and assessing electromagnetic compatibility will enable the collaboration among applications. Many military systems today are developed in isolation from other systems, or at most with well-defined interfaces to other systems. Enabling sharing of major portions of their hardware while meeting their performance requirements will be challenging. Common array apertures, waveforms, and cognitive algorithms that support the requirements of all applications will be needed. Figure 2 shows the components that would be use in a VEMO system including requires and policies, the types of algorithms to use and applications. We will next explore some of these in further depth.

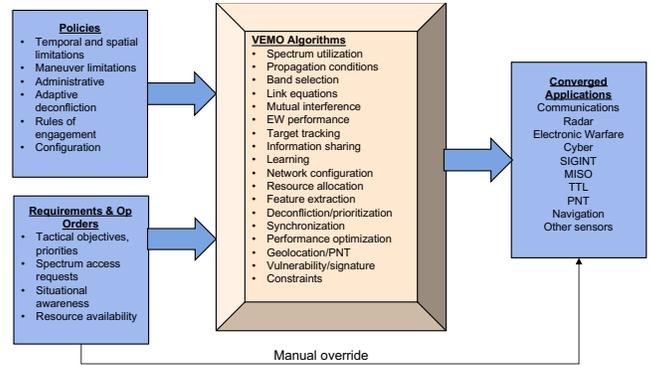

Fig. 2. Intelligent Orchestration Components

*B. Waveform Sharing*

Sharing of waveforms among applications is a key aspect of convergence. Multiplexing the application content into the signal for transmission is a primary challenge, while extracting the resulting information at a receiver is perhaps even more challenging. A trade-off between orthogonal frequency division multiplexing (OFDM) and non-orthogonal multiple access (NOMA) signaling must be considered. OFDM is spectrally efficient, and robust. However, it requires a high peak to average power ratio which imposes a limitation on the transmitter power. The use of OFDM in multiple-input, multiple-output (MIMO) communications makes it a good choice when starting with communications applications. MIMO radar is also possible in which each transmitting antenna element uses a unique signal. Wideband pulse doppler radar will be supported by OFDM such that subcarriers can be chosen with desired spacing and time-bandwidth products. It may be possible to allocate OFDM subcarriers to differing applications on a dynamic basis, much like Long Term Evolution (LTE) allocates sub-carriers to different users. Likewise, multi-band radars will be able to employ either OFDM or NOMA in waveform sharing. However, the applicability of MIMO to multi-function surveillance, tracking, and fire control radar requires study. NOMA has a good characteristic for combining applications waveforms in that a high-power signal such as EW or radar can be employed as an outer signal. Beyond that a communications signal can be obtained by successive interference cancellation (SIC) of the outer higher power signal, uncovering the communications signal. NOMA is useful in such cases where a series of receivers are at differing ranges and thus signals destined to each are non-orthogonally added at the transmitter and peeled back at each receiver, given knowledge of the signals to be removed. This can be used to improve dynamic range requirements of the transceiver, and it eliminates the peak-to-average power issue. Non-cooperative receivers, such as radar or EW targets will not perform SIC, and thus must be addressed using the primary (highest power) NOMA signal. SIC is also key to detecting signals that are below the noise floor since higher power signals can be cancelled, allowing cyclostationary processing to detect the remaining hidden signals.

The role of EW is to disrupt adversary sensing and communications and to hide friendly communications from an adversary. Combining an EW signal with communications may be achieved by the ability to cancel the EW signal at the intended communication receiver, without enabling an adversary to do the same. The use of a transmission security (TRANSEC) based EW signal may enable such cancellation. The intended receiver, holding the TRANSEC key, is able to use it to cancel the jamming, whereas an eavesdropper cannot do the same. Combing EW with radar is problematic in that the EW signal may be of the form that maximizes detrimental impact upon an adversary system and thus is designed based upon the targeted receiver. Whereas the radar signal has properties such as pulse repetition frequency that are selected based on range to the target. An OFDM signal in which subcarriers are allocated to the various applications may be a good solution. A compressive sensing approach may also be possible in which the combined signal is composed of sparse orthogonal components. Machine learning techniques such as principal component analysis may be useful in separating those signals. A benefit of the combination of the radar and EW signals is the obfuscation of the radar signal itself by the EW, thus disguising the presence of the radar. These methods are also suited for multiple aperture realizations such as multi-static radar or radar signal relay by intelligent surfaces. Ultimately the radar sensing may help inform and compose the nature of the EW attack. Sub-Nyquist compressive sensing can also be used to extend the coverage of spectrum sensors from 100's MHz instantaneous bandwidth to 10's of GHz sensing bandwidths in a second.

*C. Platform SWAP*

The VEMO concept is expected to be relevant across a range of platforms from ships to aircraft to small form factor platforms. There will be different types of options on small SWAP platforms, such as unmanned aerial vehicles (UAVs), unmanned undersea vehicles (UUVs), autonomous ground vehicles, unmanned surface ships, and low earth orbiting (LEO) satellite payloads. Larger general purpose processors may not be suitable but other FPGA-based virtualization approaches could be used. Overall, such small platforms benefit greatly from small SWAP due their extreme limitations on space and power. Virtualization of applications on small platforms will enable the deployment of added capabilities that may not be otherwise achievable on a single platform. It is well recognized that large surface combatants, tactical ground platforms and aircraft have major SWAP limitations above and below deck due to the proliferation of independent systems. SWAP is always an important consideration in commercial devices as well as consumers want compact devices.

The platform SWAP will drive the arrays that are possible to support the VEMO concept. Antennas have been considered one limitation toward the full extent of flexibility that can be achieved with a virtualized set of higher layer protocols. However, there are approaches that can combine the cognitive radio protocol layers with the adaptation that can be achieved at the antenna layer which can have benefits for mitigations against threats [11]. These concepts could be extended to other RF functions like radar. The use of a shared array antenna opens up additional concept with VEMO. A digital phased array would be conducive to the VEMO concept as it would allow addition and transmission of multiple signals. The signals can be independently phased to achieve a desired direction of transmission and reception. Beamforming can be used to improve gain while rejecting interference from undesired VEMO signals sharing the band and array. MIMO techniques could be used to support radar and communications.

Recent advances in array technologies make collaboration among applications feasible and offer some new capabilities. The advantage of digital arrays in combining signals and providing flexibility in beamforming are recognized. However, the cost and complexity of the full digital array are substantial. Two emerging concepts for array augmentation are worthy of examining in the context of convergence; holographic antennas and intelligent reflecting surfaces (IRS). Holographic apertures significantly reduce the complexity of a digital array by using reduced processing at each element. The IRS also simplifies the aperture by acting as a passive relay between a transmitter and receiver, which can be used to improve end-to-end SNR in difficult propagation environments. The IRS has no transmitter or receiver capability. However, it is capable of beamforming, managed by a low powered controller. In sensing, the IRS can employ massive-MIMO beamforming techniques to accumulate the scattered MIMO signals as a base station would, subject to the challenge of estimating channel state dynamically, for which there are techniques in research. The IRS may be useful in small SWAP cases where multiple platforms are present or in larger platforms when positioned apart from the primary antenna. For instance, the IRS can improve coverage areas of DoD comms, allow radar to look around corners, or focus EW and surveillance staring.

*D. Multi-Platform Approaches and Resource Scheduling*

The VEMO concept not only applies to individual platforms but is necessarily used across multiple platforms to orchestrate functions. This is clearly the case with SDRs where both transmit and receive nodes must be synchronized with respect to protocols to be able to communicate. There will be some functions where this internode coordination is not necessary. For example, a single platform could be optimized with respect to its resources for radar functionality or jamming. But the VEMO concept opens up the potential for coordinated platform activities like bi-static radar or multi-node jamming prosecution and in those cases coordination is necessary. As depicted in Fig 3, there will need to be some type of control plane that helps orchestrate the functionality across these individual nodes (Ax for aircraft x, Tx for tank x, EAx for enemy aircraft x and ETx for enemy tank x). This connectivity would in itself be another requirement that is levied on the VEMO system, likely with increased priority given its role. In a high threat environment it may be necessary to lower data rates to enhance robustness or plan for temporal outages by pre-scheduling VEMO-based operations. This control plane is a fruitful area for research.

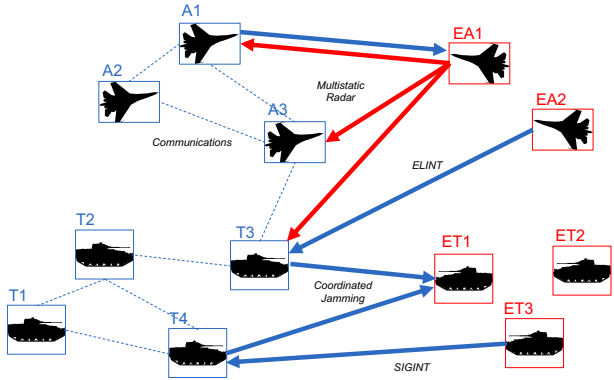

Fig. 3. Multi-Platform Control Plane

One of the important aspects to the realization of the VEMO vision is the concept of scheduling. Scheduling will be necessary to orchestrate the activities of the different RF missions to manage overlaps appropriately. This scheduling problem could leverage numerous approaches in optimization like constrained linear programming, job shop scheduling and others. Depending on the scale of operations and the potential for complexity challenges, there may be modern artificial intelligence approaches to learn behaviors though these would need to be performed with recognition of assured performance and protection against threats.

We illustrate a simple example of the resource scheduling problem based on the actions shown in Fig. 3. To keep this extremely rudimentary we assume all activities are in the same frequency band and that each platform has a single directional aperture and we do not consider power issues. Our main degree of freedom is therefore temporal scheduling of activities, we assume over discrete epochs, and an example schedule is shown in Fig 4. There are pairs of activities for communications between nodes which include both the military information messages plus the VEMO control orchestration traffic which is not differentiated per earlier discussions. The use of multi-platform approaches is shown with A1 sending a radar pulse which is received by A1, A3 and T3; however, these will likely be at different epochs given spatial differences between platforms and signal travel time. VEMO would need to be cognizant of those range differences. We see a similar situation with coordinated jamming with T3 and T4 sending jamming pulses at epochs 2 and 3 to attack ET1. Receive only operations like SIGINT and ELINT could be worked into the schedule more flexibly unless some a priori knowledge indicated when those receptions would be most valuable. VEMO would need to be fed by upstream requirements from the users like platforms that are critical to track (e.g., EA1) which then lead autonomously to the VEMO decisions about radar and associated platforms. Similarly, with those platforms targeted for jamming and forms of intercept. The human operators only enter high-level requirements and VEMO must orchestrate the rest. Based on this extremely simple case, we can imagine the complexity, and therefore further R&D, on the scheduling which includes all attributes and degrees of freedom.

| | Epoch 1 | Epoch 2 | Epoch 3 | Epoch 3 | Epoch 4 | Epoch 5 | ... |
|---|---|---|---|---|---|---|---|
| A1 | Tx radar pulse | | Comm Tx to A2 | Comm Rx from A2 | | | ... |
| A2 | Comm Tx to A3 | Rx radar pulse off EA1 | Comm Rx from A1 | Comm Tx to A1 | Comm Tx to A3 | | ... |
| A3 | Comm Rx from A2 | Rx radar pulse off EA1 | Comm Tx to T3 | | Comm Rx from A2 | Comm Rx from T3 | ... |
| T1 | Comm Tx to T2 | Comm Rx from T2 | | Comm Tx to T4 | Comm Rx from T4 | | ... |
| T2 | Comm Rx from T1 | Comm Tx to T1 | | | Comm Tx to T3 | | ... |
| T3 | ELINT Rx from EA2 | Jam Tx to ET1 | Comm Rx from A3 | Rx radar pulse off EA1 | Comm Rx from T2 | Comm Tx to A3 | ... |
| T4 | | SIGINT Rx from ET3 | Jam Tx to ET1 | Comm Rx from T1 | Comm Tx to T1 | | ... |

Fig. 4. Resource Scheduling

Much more advanced approaches to scheduling can be created. Through the use of compression and sparsity techniques it may be possible to excise portions of a communications or radar signal to allow use of the channel for passive sensing or for EW. Likewise, EW may be temporarily halted to allow communications or radar sensing. Puncturing of error correction codes can make resources available or use of strong error correction can tolerate excision of portions of a signal without any significant detrimental impact.

*E. VEMO Example: Signal Combining*

In this section we expand on a particular benefit of VEMO which is the use of NOMA or OFDM to combine signals together which opens a new dimension beyond the temporal scheduling approach discussed in the previous section. The signals used by differing applications can be combined in a non-orthogonal manner or in an orthogonal manner.

Consider a radar signal that can be transmitted and provide a return from a target. At the same time, the same signal can carry a communications message to a desired cooperative receiver. Further, the same signal can be used to intentionally interfere with another non-cooperative receiver. Hence, we desire to wrap a communications signal inside a radar signal, and also to allow the resulting signal to be used for electronic warfare as well. The radar signal is emanated by a transmitter, impinges on a target, and is reflected back to a receiver (collocated) with the transmitter. The communications signal is transmitted by the same transmitter as the radar signal, but it is received by a destination receiver. The EW signal is transmitted by the same transmitter, but is received by a victim receiver that is to be jammed. The intended victim is receiving a signal from another intended transmitter.

In this arrangement, the radar target is not cooperative in the sense that it does not intentionally process the radar signal. Of course, radar targets do passively and can actively alter the radar signal. The communication receiver is able to perform signal processing on the received signal, and that processing can be designed to support the VEMO processing cooperatively. The intended jamming victim may processing the incoming VEMO signal with the goal of suppressing it, while improving the signal it receives from the intended transmitter. Thus, the combined VEMO signal is processed differently at each target/receiver/victim at which it is received or reflected. The radar and EW targets are noncooperative, while the communications receiver is cooperative.

The VEMO system may also be supporting sensing at the same time as the above. In sensing we do not control the emitted signal, but we do select the processing to be done. We must not transmit in the same frequency band while receiving. The radar plus comms signal is transmitted. The communications receiver can demodulate the strong radar signal and subtract it from the received signal using SIC, leaving a version of the communications signal. The receiver then demodulates the remaining communications signal. Concurrently the radar signal reflects from the target and returns to the tower as a radar return. Design of the system will have to account for imperfect cancellation and multipath in the propagation environment. When the transmitter employs an antenna array, beamforming of the radar signal is possible, and MIMO transmission of the communications signal is also possible, concurrently as a combined signal.

VEMO signals also can be combined using OFDM. The subcarriers can compose radar and communication signals which will result in various performance impacts for each of the system's functionality.

V. SUMMARY AND FUTURE WORK

The VEMO concept could fundamentally change systems that host multiple types of RF functionality. Current systems are stove piped and lack the ability to consider more cohesive and efficient spectrum usage across applications. We have provide the framework and some examples for VEMO as well as pointed out some of the challenges.

We have noted several areas that warrant further technical investigation throughout the paper. The ability to develop the architecture of virtualized operations with the different types of processing capabilities will be important from individual capabilities and the broader use. There is a wide open research space in the area of techniques for simultaneous and coordinated mission functionality. Since many of these fields have worked independently for decades, these cross-cutting ideas that exploit relationships between communications, radar, jammers and other have not been explored. We have touched on one of the limitations of physics which is in the area of apertures and RF components that have limits on dynamic range. However, this should not hold back the significant potential that is offered using phased array antennas even confined to particular bands. These would be introduced as new constraints in the algorithms that orchestrate VEMO. Optimization techniques at machine speeds are one of the most important for VEMO. These autonomous algorithms need to support optimization of large-scale, complex and multi-dimensional EM problems. Any one of these attributes makes this a problem beyond what a human operator could achieve. Finally, for military operations, we need to ensure some level of security of virtualized operations. This puts significant functionality into one system so the ability to assure the operations of the software components of VEMO and the larger functionality is critical.

It is important to note that there would also be regulatory and policy imperatives that relate to spectrum usage. Just as cognitive radio has been working through these through primary and secondary allocation concepts, expanding VEMO to include other RF will require new ways to assign and manage spectrum.